\documentclass[twocolumn,showpacs,amsmath,amssymb,prd,nofootinbib]
{revtex4}
\usepackage{graphicx}
\def\sss{\scriptscriptstyle}
\def\^#1{^{\sss #1}}
\def\_#1{_{\sss #1}}
\def\beq{\begin{equation}}
\def\eeqno#1{\label{#1}\end{equation}}

\def\rarrow{\rightarrow }

\def\dleft{\rlap{{\it D}}\raise 8pt\hbox{$\scriptscriptstyle\Leftarrow$}}
\def\dright{\rlap{{\it
D}}\raise 8pt\hbox{$\scriptscriptstyle\Rightarrow$}}

\def\cmss{{\rm cm~s^{-2}}}
\def\kpc{{\rm ~kpc}}

\def\msun{M\_{\odot}}
\def\az{a\_{0}}

\def\l0{\ell\_{0}}

\def\vinf{V_{\infty}}

\def\l{\lambda}

\def\m{\mu}
\def\n{\nu}

\def\a{\alpha}

\def\xlimin{{x\rarrow\infty \atop{\raise 1pt\hbox to 30pt{\rightarrowfill}}}}
\def\limlim#1#2{{#1\rarrow #2 \atop{\raise 1pt\hbox to 30pt{\rightarrowfill}}}}

\def\gN{g\_N}

\def\zef{\zeta\_{1/2}}
\def\raf{R\_{1/2}}

\newcommand{\kms}{\ensuremath{\mathrm{km}\,\mathrm{s}^{-1}}}

\begin{document}
\title{High-redshift rotation curves and MOND}
\author{Mordehai Milgrom} \affiliation{Department of Particle Physics and Astrophysics, Weizmann Institute of
Science, Rehovot 76100, Israel}

\begin{abstract}
Genzel et al. have recently published the rotation curves (RCs) of six high-redshift disc galaxies ($z\sim 0.9-2.4$), which they find to be `baryon  dominated' within the studies radii. While not up to the standard afforded by data available for analysis in the nearby Universe, these data are valuable in constraining cosmological evolution of either DM scenarios, or -- as I discuss here -- $z$-dependence of MOND. Indeed, these results, if taken at face value, teach us useful lessons in connection with MOND.
a. The dynamical accelerations at the half-light radii, found by Genzel et al., are rather high compared with the MOND acceleration constant, as measured in the nearby Universe:
$g(\raf)= (3-11)\az$. MOND then predicts fractions of `phantom matter' at $\raf$ of at most a few tens of percents, which, galaxy by galaxy, agree well with what Genzel et al. find. b. The asymptotic rotational speeds predicted by MOND from the baryonic-mass estimates of Genzel et al. are substantially lower ($0.55-0.75$) than the maximal speeds of the RCs.
MOND thus predicts a substantial decline of the RCs beyond the maximum. This too is in line with what Genzel et al. find. c. Arguably, the most important lesson is that the findings of Genzel et al. cast very meaningful constraints on possible variation of $\az$ with cosmic time. For example, they all but exclude a value of the MOND constant of $\sim 4\az$ at $z\sim 2$, excluding, e.g., $\az\propto (1+z)^{3/2}$.
\end{abstract}
%\keywords{}
\maketitle

\section{introduction}
\label{introduction}
By far the most acute and clear-cut tests of MOND \cite{milgrom83a} come from the dynamical analysis of rotation curves (RCs) of disc galaxies. For reviews of MOND see, e.g. Refs. \cite{fm12,milgrom14}.
\par
In particular, the MOND acceleration constant, $\az$, appears in such analyses in different roles: as the boundary constant marking the transition from Newtonian behavior to deep-MOND, as setting behavior in the deep MOND regime, for example, fixing the normalization of the MOND mass-asymptotic-speed relation (MASR) -- underlying the baryonic Tully-Fisher relation (BTFR), and in dictating the RCs of low-acceleration (or low-surface-brightness) galaxies. It appears in several roles in the much discussed MOND prediction \cite{milgrom83a} of the mass-discrepancy-acceleration relation (for tests of this prediction see, e.g., \cite{sanders90,mcgaugh04,tiret09,wu15,mcgaugh16a}). It also appears in the no-less-striking central-surface-densities relation, which is different and independent of the other MOND relations \cite{milgrom09a,lelli16,milgrom16a}.
\par
The value that has emerged for $\az\approx 1.2\times 10^{-8}\cmss$, as been recognized early on (Milgrom 1983a) to have cosmological connotations. In particular we have:
\beq \bar\az\equiv 2\pi \az\approx cH_0\approx c^2(\Lambda/3)^{1/2}, \eeqno{coinc}
where $H_0$ is the Hubble constant, and $\Lambda$ the observed equivalent of a cosmological constant.
\par
The Former of these near equalities, and the realization that MOND may well be an effective theory rooted somehow in cosmology, have pointed to the possibility that $\az$, or some aspects of MOND, may be varying with cosmological time  so as to retain the first equality at all times.

The obvious way to test this possibility, given that $\az$ is sharply determined by rotation-curve analysis, is to analyze RCs of high-$z$ galaxies to see if their dynamics can be accounted for by MOND, and whether this requires $\az$ to be $z$ dependent (an early attempt at this is described in Ref. \cite{milgrom08}).
\par
In recent years, there have been several studies of the internal kinematics of high-$z$ galaxies (e.g., \cite{price16,wuyts16,lang17}). These are, by and large, statistical in nature.
\par
Genzel et al. \cite{genzel17} have recently published the individual RCs of six high-redshift galaxies
($z\sim 0.9-2.4$), and have presented a thorough dynamical analysis of them.
These are selected from a large sample of several hundred, according to criteria that are conducive to cleaner analysis.
This sample now also affords a closer, if preliminary, examination of the dynamics of high-$z$ in light of MOND.
\par
The main general conclusions of Ref. \cite{genzel17} are that these galaxies are `baryon dominated' within the studies radii, and that they show marked decline in the RC still within the optical image.
In both regards, this is very reminiscent of the findings of Ref. \cite{romanowsky03} of `dearth of dark matter in ordinary elliptical galaxies' (at low redshift) -- based on planetary-nebulae velocities.
\par
I shall show that both of these characteristics of the high-$z$ disc galaxies of Ref. \cite{genzel17} follow from MOND because these galaxies have accelerations within the studied regions that are higher than $\az$. The MOND analysis by Ref. \cite{ms03} and Ref. \cite{tian16} showed this for low-$z$, elliptical galaxies.
\par
It is important to keep in mind that for natural reasons the data of Ref. \cite{genzel17} are not up to the standard afforded by RCs and baryon distributions available for dynamical analysis in the nearby Universe. In comparison with the latter they are limited in scope, and they are subject to large uncertainties (partly reflected in their large quoted errors).
Some concerns that come to mind are: a. The inclinations of the six galaxies are $i({\rm deg})=75\pm 5,~30\pm 5,~62\pm 5,~25\pm 12,~45\pm 10,~34\pm 5$. Three of them have low inclinations $i< 35$ degrees. Such low inclinations are generally considered problematic because it is difficult to measure such low inclinations accurately, and because the actual rotational speeds, and the acceleration deduced from them are sensitive to the exact value (the accelerations scaling as $1/sin^2 i$).
b. This is further compounded by the fact that
these RCs are not based on 2-D tilted-ring derivation as the standard has come to require, and hence do not account for possible variable position angle and inclination, especially problematic for low-inclination galaxies.
c. Large random motions are present in these galaxies; so large (and uncertain) asymmetric-drift corrections have to be applied. d. The luminosity distribution is measured in the rest-frame optical-band, not as good for converting light to mass compared with far IR now used routinely for local galaxies. e. Some of the galaxies have a substantial bulge, and the necessary separation to components, with possibly different M/L values, is problematic.
f. Kinematics are measured from $H_\a$ velocities, so are confined to the optical image with no analog of the extended HI RCs.\footnote{In many of these regards the quality of these RCs may be likened to that of RCs available in the late 1970s for low-$z$ galaxies, before extended HI RCs became available.}
\par
Still, these are the best RC data we have at present for such high redshift, and thus are valuable in constraining cosmological evolution of either DM scenarios, or, as here, $z$-dependence of MOND.
\par
In Sec. \ref{MOND}, I give some MOND formulae needed here. Section \ref{results}
compare the MOND predictions with the results of Ref. \cite{genzel17},
and Sec. \ref{discussion} is a discussion.

\section{Relevant MOND formulae}
\label{MOND}
If at some radius, $R$, in the midplane of a disc galaxy $\gN(R)$ is the Newtonian acceleration calculated from the baryon distribution, and $g$ is the dynamically determined acceleration, then MOND predicts the relation \cite{milgrom83a}:
\beq g\m(g/\az)=\gN, \eeqno{mdarmu}
where $\az$ is the MOND acceleration constant, and $\m(x)$ the `interpolating function' for rotation curves.
This can be written equivalently in terms of the $\n(y)$ interpolating function
as
\beq g=\gN\n(\gN/\az), \eeqno{mdarnu}
where $\n(y)$ is related to $\m(x)$  by $\m(x)=1/\n(y)$, where $x=y\n(y)$ [and so $y=x\m(x)$].
These equivalent forms are known as the mass-discrepancy-acceleration relation (MDAR) since
\beq \eta\equiv g/\gN=1/\m(g/\az)=\n(\gN/\az) \eeqno{mdar}
can be identified as the mass discrepancy.
\par
As convention goes, Ref. \cite{genzel17} define the dark-matter fraction at $R$ -- better referred to in the present context as the `phantom-matter' fraction --as
$\zeta(R)=[V\_{DM}(R)/V(R)]^2$, or in terms of the accelerations
\beq \zeta(R)\equiv (g-\gN)/g. \eeqno{frac}
Thus, MOND predicts
\beq \zeta=1-\eta^{-1}=1-\m=1-1/\n.  \eeqno{etamu}

I will show the MOND predictions for $\zeta$ for two forms of the MOND interpolating function used routinely for RC fits and fits to the MDAR.
The first is
\beq \m(x)=\frac{x}{1+x}, \eeqno{i}
which gives a `phantom-matter' fraction of $\zeta=(1+x)^{-1}$.
The other takes a simple form in the $\n(y)$ language \cite{ms08,mcgaugh08,mls16,milgrom16}:
\beq \n(y)=(1-e^{-\sqrt{y}})^{-1},   \eeqno{ii}
which gives $\zeta=e^{-\sqrt{y}}$.
These two interpolating functions differ by at most $\sim 5\%$ over the full range of arguments and so predict almost indistinguishable rotation curves.
However, in the region of high accelerations, where both functions are nearly 1, they differ substantially in their exact departure from 1. So, when the predicted
fractions of `phantom matter' are small, the two functions can give different predictions for this small quantity.
\par
Another MOND prediction I will need: Given the total baryonic mass, $M_b$, MOND predicts \cite{milgrom83a,milgrom83b} for an isolated galaxy, an asymptotically flat RC, with the constant rotation speed
\beq \vinf^4=M_bG\az.  \eeqno{iii}
This is the MASR mentioned in Sec. \ref{introduction}
\section{Results}
%\begin{center}
\begin{table*}[t]
 \begin{tabular}{lccccccccl}
 \hline\hline

Galaxy& $z$ & $\raf$ & $V_c(\raf)$ & $M_b$& $\zef$& $\vinf$ & $x\_{1/2}$& $\zeta\^M\_{1/2,a}$ & $\zeta\^M\_{1/2,b}$  \\
& & $\kpc$ & $\kms$ & $10^{11}\msun$ & & $\kms$&&&\\
\hline
COS4 01351   &  0.854 & 7.3  & 276 & 1.7 & $0.21 (\pm0.1$) & 228 & 2.8 &0.26& 0.22 \\
D3a 6397 & 1.500&  7.4      & 310 & 2.3  & 0.17 ($<$0.38)	& 246 & 3.5 &0.22& 0.18 \\
GS4 43501  & 1.613 &  4.9  & 257 & 1.0 & $0.19 (\pm0.09)$ & 200 & 3.6 & 0.22 & 0.17 \\
zC 406690 &2.196	&  5.5   & 301  &  1.7  & $0 (<0.08)$ & 228 & 4.4 & 0.18 & 0.14 \\
zC 400569 & 2.242  &  3.3      & 364    	 & 1.7 & $0 (<0.07)$ & 228 & 10.8 & 0.08 & 0.04 \\
D3a 15504 & 2.383  &  6      & 299    	 & 2.1 & $0.12 (<0.26)$ & 240 & 4.0 & 0.20& 0.16\\
\hline
\end{tabular}
\caption{Galaxy name \{column 1\}, its redshift \{2\}. Columns 3-6 are best-fit attributes deduced by Ref. \cite{genzel17}: the half-light radius, $\raf$, (in the rest-frame optical band) \{3\}; the rotational speed there (corrected for inclination and asymmetric drift) \{4\}; the total baryonic mass \{5\}; and the dark-matter fraction, $\zef$ at $\raf$, with errors or upper limits \{6\}. Column 7-10 show calculated MOND quantities: the predicted asymptotic rotational speed, $\vinf$, based on $M_b$, from eq.(\ref{iii}) \{7\}, The acceleration at $\raf$ in units of $\az$ \{8\}, the expected MOND value of $\zef$ based on the interpolating function of eq. (\ref{i}), $\zeta\^M\_{1/2,a}$ \{9\}, and that based on eq. (\ref{ii}), $\zeta\^M\_{1/2,b}$ \{10\}, all calculated for the nearby-Universe value of $\az=1.2\times 10^{-8}\cmss$.}
\label{table}
\end{table*}
%\end{center}
\label{results}
Table \ref{table} shows the values of the relevant parameters as they appear in Table 1 of Ref. \cite{genzel17}. I show in the table, and use, the relevant quantities given in Ref. \cite{genzel17} as their best fit model parameters (resulting from fitting the rotation curves to mass models that include baryons and dark matter): the half light radius, $\raf$, the dynamical rotational speed at $\raf$, and the total baryonic mass, $M_b$. For $M_b$ and $\raf$ they also give their pre-fit, directly estimated values. In most cases, the former values, which I use, lie within the error range of the latter. I also show their deduced values of the `phantom-matter' fractions, $\zef$, at $\raf$, and the values of $\zef$ predicted by MOND for the two commonly used interpolating functions, all as detailed in Sec. \ref{MOND}.
\par
We see that as found by Ref. \cite{genzel17} the MOND $\zef$ values are small -- a few tens of percents at most. Furthermore, except for the rogue zC 406690, where the upper limit is lower than my estimates, the MOND predictions are, case by case, in good agreement with what Ref. \cite{genzel17} give.\footnote{But beware that the $\zef$ values of Ref. \cite{genzel17} are based on model best-fits with NFW dark-matter distributions. Given their large uncertainties on $M_b$, their RCs are probably also consistent with sub-maximal discs, and rather larger values of $\zef$. The distinction between maximal and sub-maximal discs is moot even with much better data.\label{fnq}} And note that zC 406690 has a quoted inclination of $i=25\pm 12$ degrees; so it's kinematic analysis is practically useless.

\subsection{Falling rotation curves}
The RCs shown by Ref. \cite{genzel17} exhibit some decline beyond their maximum. Such decline is also typical of high-surface-brightness galaxies in the local Universe (see, e.g. some early-type galaxies in the sample of Ref. \cite{sn07}, in particular, their RC for UGC 4458, which drops from $\sim 500\kms$ to $\sim 300\kms$ within $10 \kpc$ and then becomes flat at $\sim 250\kms$ to $55\kpc$).
\par
Such a decline seems to be more prevalent in the high-$z$ samples at hand (see also Ref. \cite{lang17}). Part of the reason, as extensively discussed by Refs. \cite{genzel17,lang17}, is that rather more than in low-$z$ galaxies, velocity dispersions in the disc contribute substantially to the balance against gravity, hence diminishing the role of rotational support. It is notoriously difficult and uncertain to correct for this important effect. Indeed, in the stacked RCs of Ref. \cite{lang17} (see their Fig. 8), galaxies with
high rotation-to-dispersion ratio show much less marked decline than those with small values.
\par
One should also consider the effects of selection: High-surface-brightness galaxies -- where such declines are also observed at low-$z$ -- are naturally more amenable to measurements at high redshift, and are easier to follow to larger radii. Indeed, Fig. 5 of Ref. \cite{lang17} shows that the number of galaxies contributing at the outer radii, where the decline is evident, is much smaller than the total in the sample: $\sim 12$ galaxies that contribute down to the outer stacked-data point, compared with $\sim 90$ that contribute at low radii. These may well be selecting preferentially higher-surface-brightness galaxies.
\par
In MOND, we do expect marked decline beyond the maximum in galaxies with mean accelerations that are so high compared with $\az$. For example, MOND-predicted rotation curves of such model galaxies are shown in Figs. 1 and 2
of Ref. \cite{milgrom83b} (the models with high $\xi\sim 5$ there). And see also Fig. 2 of Ref. \cite{ms03} for the predicted MOND RC of the elliptical NGC 3379, which drops from $\sim 300\kms$ at maximum to $\sim 200 \kms$.
\par
We can estimate the room for a drop in the velocity allowed by MOND for the six galaxies under study, by comparing the observed maximum speed with the predicted asymptotic rotational speed, $\vinf$, which can be deduced from the estimates of the baryonic masses, using eq. (\ref{iii}) -- assuming that the galaxy is isolated. These estimates are given in Table \ref{table} based on the best-fit values that Ref. \cite{genzel17} give for $M_b$. Note that the direct estimates of $M_b$ given by Ref. \cite{genzel17} have large quoted errors given in all cases as $\pm 50\%$ (i.e., a factor of $\sim 3$ in range), corresponding to a relative error of $+0.1~-0.15$ in $\vinf$. From Fig. 2 of Ref. \cite{genzel17} one sees that the maximum speed for the galaxies is about
$1.1V(\raf)$, and occurs at $\sim 1.5 \raf$. We see then that the estimated ratio $\vinf/V_{max}$ is as low as $\sim 0.55~ (\pm 0.1)$ (for one of the 6 galaxies, zC 400569), and is $\sim 0.7 (\pm 0.1)$ for most others. This would allow the drops Ref. \cite{genzel17} estimate (these are subject to substantial uncertainties due to the uncertain asymmetric-drift correction, and possible unaccounted for warps). For one galaxy, zC 406690, Ref. \cite{genzel17} estimate a very large drop. But, as I pointed out above, this is quite unreliable as the stated inclination for this galaxy is $i=25\pm 12$ degrees.
\par
In MOND, the presence of neighboring bodies can also contribute to the decline of the RCs through the external-field effect (e.g., Refs. \cite{milgrom83a,wu15,haghi16,mcgaugh16}. According to Ref. \cite{genzel17} their 6 galaxies are relatively isolated, so this should not be a factor, but it is hard to asses the importance of the effect in statistical studies such as that of Ref. \cite{lang17}.

\section{Discussion}
\label{discussion}
The results of Ref. \cite{genzel17} are well accounted for by MOND in the very form that has been applied successfully to low-$z$ galaxies, with the canonical value of $\az$.
\par
Although these RCs do not probe the deep MOND regime -- where MOND enters in full glory -- they do vindicate an important prediction of MOND that does not arise naturally in the dark-matter paradigm. Namely, that mass anomalies should be small (sub-dominance of `phantom matter') at accelerations above $\az$.
That this is now seen to be the case also at high $z$ even sharpens the case for MOND: It shows this prediction to be independent of the evolutionary status of the galaxies, strengthening the case for a law of nature as the origin, rather than some complicated and contrived evolutionary processes.
\par
It appears that these results cannot accommodate much higher values of $\az$ at high redshift. Looking at Table \ref{table}, we see that, for example, a value of the MOND acceleration of $4\az$ would have resulted in $x\_{1/2}$ values for the higher-$z$ galaxies of order 1. This would have predicted $\zef$ values of order 0.5, which would be uncomfortably in tension with the values estimated by Ref. \cite{genzel17}\footnote{Values of the MOND constant smaller than $\az$ cannot be excluded, but they are anyhow less motivated.} (But, remember footnote \ref{fnq}.) This constraint makes use, essentially of the role of $\az$ in MOND as `boundary acceleration'. Another, independent constraint is based on the role of $\az$ as setting the MASR normalization: With a value of the MOND constant as high as $4\az$ the predicted values of $\vinf$ in  Table \ref{table} should be increased by a factor of $4^{1/4}\sim 1.4$, making $\vinf/V_{max}\sim 1$, not leaving room for decline beyond the maximum, unless the baryonic masses are substantially lower.
\par
Ideally, we could test for variations of $\az$ by searching for evolution in the proportionality constant of the MASR, eq. (\ref{iii}). But this is not possible with the present data, as clearly they do not reach the asymptotic speeds, as required by the MOND MASR.  `Evolution' of the zero point of some versions of the BTFR, using available velocity measures such as the maximum speed have been studied. But these are not what the MOND MASR dictates, and cannot be used to constrain cosmological variations of the MOND constant.
It is an opportunity to stress again the distinction between various versions of the BTFR, and the specific version MOND predicts as the MASR, which employs the asymptotic speed.
\par
This result may help constrain ideas that rest on the MOND constant varying with cosmic time, such as the suggestion that the first of the near equalities in eq. (\ref{coinc}) held at all times, or other possible variations (see discussion in Ref. \cite{milgrom09} and references therein). Ref. \cite{milgrom99} offers a possible causal connection between $\az$ and $\Lambda$.

\clearpage
\end{document}